\documentclass[10pt, conference]{IEEEtran}
\usepackage{cite}
\usepackage{amsmath,amssymb,amsfonts}
\usepackage{algorithmic}
\usepackage{graphicx}
\usepackage{textcomp}
\usepackage{xcolor}
\usepackage{xspace}
\usepackage{url}
\usepackage{tcolorbox}
\usepackage{booktabs}
\usepackage{enumitem}

\newcommand{\aim}{{\em Assertion Inferring Mutants}\xspace}
\newcommand{\our}{{\em AIMS}\xspace}
\newcommand{\sms}{{\em Subsuming Mutant Selection}\xspace}
\newcommand{\rnd}{{\em Random Mutant Selection}\xspace}

\makeatletter
\newcommand{\linebreakand}{%
  \end{@IEEEauthorhalign}
  \hfill\mbox{}\par
  \mbox{}\hfill\begin{@IEEEauthorhalign}
}
\makeatother

\title{Assertion Inferring Mutants}

\author{
\IEEEauthorblockN{Aayush Garg}
\IEEEauthorblockA{\textit{University of Luxembourg} \\
\textit{Luxembourg} \\
aayush.garg@uni.lu}
\and
\IEEEauthorblockN{Renzo Degiovanni}
\IEEEauthorblockA{\textit{University of Luxembourg} \\
\textit{Luxembourg} \\
renzo.degiovanni@uni.lu}
\and
\IEEEauthorblockN{Facundo Molina}
\IEEEauthorblockA{\textit{IMDEA Software Institute} \\
\textit{Spain} \\
facundo.molina@imdea.org}
\linebreakand
\IEEEauthorblockN{Mike Papadakis}
\IEEEauthorblockA{\textit{University of Luxembourg} \\
\textit{Luxembourg} \\
michail.papadakis@uni.lu}
\and
\IEEEauthorblockN{Nazareno Aguirre}
\IEEEauthorblockA{\textit{University of Río Cuarto} \\
\textit{Argentina} \\
naguirre@dc.exa.unrc.edu.ar}
\and
\IEEEauthorblockN{Maxime Cordy}
\IEEEauthorblockA{\textit{University of Luxembourg} \\
\textit{Luxembourg} \\
maxime.cordy@uni.lu}
\and
\IEEEauthorblockN{Yves Le Traon}
\IEEEauthorblockA{\textit{University of Luxembourg} \\
\textit{Luxembourg} \\
yves.letraon@uni.lu}
}

\begin{document}

\maketitle

\begin{abstract}

Specification inference techniques aim at (automatically) inferring a set of assertions that capture the exhibited software behaviour 
by generating and 
filtering assertions through dynamic test executions 
and mutation testing. 
Although powerful, such techniques are computationally expensive 
due to a large number of assertions, test cases and mutated versions that need to be executed. To overcome this issue, we demonstrate that a small subset, i.e., 12.95\% of the mutants used by mutation testing tools is sufficient for assertion inference, this subset is significantly different, i.e., 71.59\% different from the subsuming mutant set that is frequently cited by mutation testing literature, and can be statically approximated through a learning based method. In particular, we propose  \our, an approach that selects \aim, i.e., a set of mutants that are well-suited for assertion inference, with 0.58 MCC, 0.79 Precision, and 0.49 Recall. We evaluate \our on 46 programs and demonstrate that it has comparable inference capabilities with full mutation analysis (misses 12.49\% of assertions) while significantly limiting execution cost (runs 46.29 times faster). A comparison with randomly selected sets of mutants, shows the superiority of \our by inferring 36\% more assertions while requiring approximately equal amount of execution time. We also show that \our’s inferring capabilities are almost complete as it infers 96.15\% of ground truth assertions, (i.e., a complete set of assertions that were manually constructed) while \rnd infers 19.23\% of them. More importantly, \our enables assertion inference techniques to scale on subjects where full mutation testing is prohibitively expensive and \rnd does not lead to any assertion.

\end{abstract}


\section{Introduction}
\label{sec_introduction}
Software specifications aims at describing the software's intended behavior, and can be used to distinguish the corresponding correct/expected software behaviour from the incorrect/unexpected one. 
While these are typically described informally (e.g. API documentation), specifications become significantly more useful when expressed formally, in the form of executable constraints/assertions. 
Executable specifications are typically composed of  code assertions for various program points, such as method preconditions and postconditions, that must hold true during the program execution. 
These are known to be useful in many software engineering tasks, e.g., test generation \cite{DBLP:conf/kbse/dAmorimPXME06, DBLP:conf/tap/TillmannH08}, bug finding \cite{DBLP:journals/scp/LeavensCCRC05, DBLP:conf/icse/PachecoLEB07} and automated debugging \cite{DBLP:conf/issta/DemskyEGMPR06, DBLP:conf/oopsla/LogozzoB12, DBLP:conf/sosp/PerkinsKLABCPSSSWZER09}. However, they are tedious to write and maintain, and as a result developers often avoid writing them~\cite{DBLP:conf/issta/BlasiGKGEPC18, DBLP:conf/icse/WatsonTMBP20}. 

To address this issue, specification inference techniques aim at automatically inferring assertions for specific program points that capture the exhibited software behaviour~\cite{DBLP:conf/sigsoft/TerragniJTP20, DBLP:conf/icse/MolinaPAF21, DBLP:conf/icse/MolinadA22}. 
These techniques evolve candidate assertions and use dynamic test executions to determine which of those assertions are consistent with the behaviours exhibited by a provided test suite, and mutation testing to discard ineffective/weak assertions that are unable to detect any artificially seeded fault (mutant), i.e., assertions never falsified during mutant's execution. 
Though powerful, these techniques are computationally expensive due to a large number of tests, assertions and mutant executions involved. 
The problem is further escalated when working with large programs as the number of mutants grows proportionally to the program size. For instance, state of the art technique SpecFuzzer~\cite{DBLP:conf/icse/MolinadA22} times out (requires more than 90 minutes to run) in programs with 180 lines of code. 

To reduce the computational demands, it is imperative to limit the number of mutants involved since fewer mutants yield fewer executions.
Interestingly, we find that the majority of the mutants used by existing assertion inference techniques are redundant, meaning that discarding them do not impact the quality of inferred assertions. 
We denote as \aim to the subset of mutants produced by a mutation testing tool that can be used to identify all the effective candidate assertions (i.e. those falsified during mutants' execution) that can also be identified as effective by the entire set of mutants.

We demonstrate that \aim represent 12.95\% of the mutants supported by Major~\cite{DBLP:conf/kbse/JustSK11} (the mutation testing tool employed in previous studies), allowing for drastic assertion inference overhead reductions. At the same time, \aim are significantly different from subsuming mutants (which have been studied by the literature \cite{DBLP:journals/ac/PapadakisK00TH19, 9677967}) with 71.59\% of them not being subsuming. This means that subsuming mutant selection techniques are ineffective for assertion inference, as they would miss many assertions (48.53\% according to our results).

We thus propose \our\footnote{\emph{Assertion Inferring Mutant Selector (AIMS)}}, a learning-based technique to statically identify assertion inferring mutants given their contextual information. In particular, \our learns the associations between mutants and their surrounding code with respect to the assertion inference task. This means that our learning scope is the area around the mutation point that identifies locally, the mutants that are most likely to be useful from those that are not. 

\our operates at the lexical level, with a simple code pre-processing that represents mutants and their surrounding code as vectors of tokens with all user defined identifiers (e.g. variable names) replaced by predefined and predictable identifier names. 
This representation allows us to restrict the related vocabulary and the learning scope to a relatively small fixed size of tokens around the mutation points enabling inter-project predictions. Code embeddings extracted from an encoder-decoder architecture~\cite{DBLP:conf/emnlp/KalchbrennerB13} that we train on code fragments, are extracted and learned with corresponding labels using a classifier~\cite{DBLP:journals/ml/Breiman01}.

We implement \our and evaluate its ability to predict \aim on a large set of 46 programs, composed of 40 taken from previous studies~\cite{DBLP:conf/sigsoft/TerragniJTP20, DBLP:conf/icse/MolinaPAF21, DBLP:conf/icse/MolinadA22} and 6 large Maven projects taken from GitHub to evaluate scalability. Our results demonstrate that \our can statically select \aim with 0.79 Precision and 0.49 Recall, overall yielding 0.58 MCC\footnote{\emph{Matthews Correlation Coefficient}~(MCC)~\cite{DBLP:conf/ease/YaoS20} is a reliable metric of the quality of prediction models~\cite{DBLP:journals/tse/ShepperdBH14}, relevant when the classes are of different sizes, e.g., \emph{12.95\%} \aim in total (in comparison to 87.05\% low utility mutants), for subjects in our dataset.}. At the same time, since \our selects fewer mutants than previous work, it improves assertion inference scalability allowing it to run on all the projects we considered where previous work failed.

Surprisingly, by performing assertion inference based only on \our's predicted mutants (instead of all mutants), we reduce assertion inference time (wall clock) by 46.29 times with only 12.49\% assertion missed. 
Additionally, when comparing with randomly selected sets of mutants (same number as those selected by \our), we observe a clear superiority \our in terms of effectiveness, i.e., \our infers 36\% more assertions while taking approximately equal amount of execution time as \rnd.

Finally, we show that \our{}’s inferring capabilities are almost complete as it infers 96.15\% of ground truth assertions, (i.e., the complete set of assertions that were manually validated) while \rnd infers 19.23\% of them. More importantly, \our enables assertion inference techniques to scale by allowing its operation on all 6 real-world subjects we selected, where full mutation testing is prohibitively expensive. In half of these subjects, \rnd does not lead to any assertion inference and it is subsumed by \our in the other half of the subjects. 

To sum up, our paper makes the following contributions:
\begin{enumerate}[leftmargin=*]
\item We show that effective assertion inference can be performed using only 12.95\% of the mutants. We also show that these set of assertion inferring mutants is significantly different (i.e., 71.59\% different) from the subsuming mutant set, a reference class of mutants frequently used by mutation testing literature. 
\item We propose \our, a static mutant selection technique that predicts \aim with a good performance (0.58 MCC, 0.79 Precision, and 0.49 Recall). When performing assertion inference, \our allows inferring 46.29 times faster the assertions, that could be inferred with the full set of mutants, at the expense of 12.49\% missed assertions. 
We also show that \our is significantly more effective than random and subsuming mutant selection baselines. That is, \our infers 36\% and 30\% more of the assertions, that can be inferred with the full mutant set, than random and subsuming mutant selection. 

\item  We show that \our’s inferring capabilities are almost complete as it is capable to infer 96.15\% of ground truth assertions (i.e., the complete set of assertions that were manually constructed to evaluate previous work). It should be noted that the alternative baselines approaches - \sms and \rnd infer far fewer assertions, i.e., 67.31\% and 19.23\% of the ground truth assertions, respectively. 

\item Finally, we show that \our improves the scalability of SpecFuzzer by allowing it to run in programs for which it was not able to run before. Precisely, \our allows assertion inference in cases where random selection fails (50\% of the cases we tried).
\end{enumerate}

\section{Background \& Related Work}
\label{sec_background}

\begin{figure}[t]
\centerline{\includegraphics[width=0.5\textwidth]{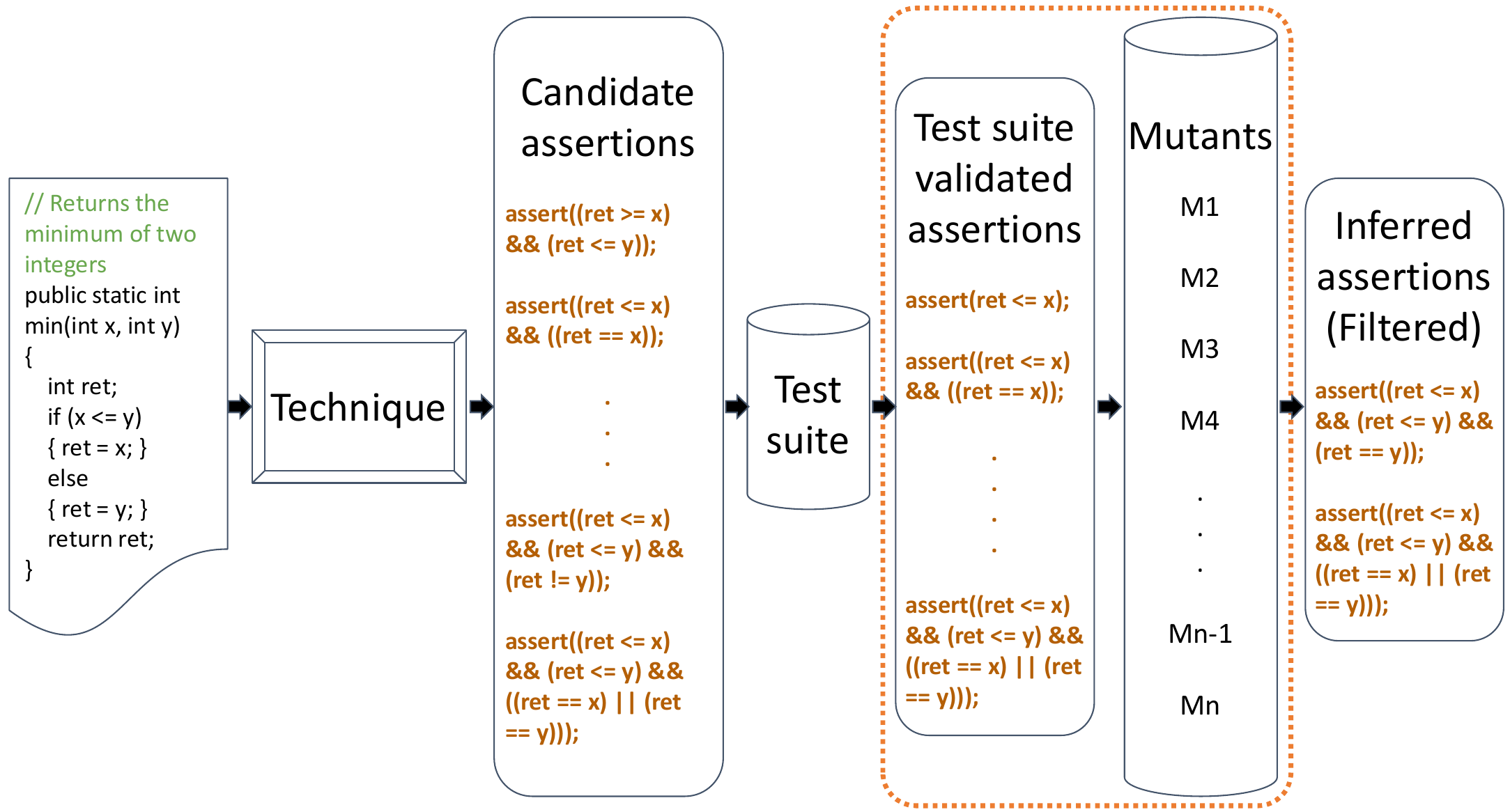}}
\caption{Assertion Inference with Filtering via Mutation Analysis}
\label{fig_assertion_inference}
\end{figure}

\subsection{Assertion Inference}
A code assertion is a logical expression capturing a property that should hold at a specific program location. It is often used as an executable description of the expected software behavior that has widespread applications in software design~\cite{DBLP:books/ph/Meyer97}, software testing~\cite{DBLP:books/daglib/0020331}, and verification~\cite{DBLP:journals/sigsoft/ClarkeR06,DBLP:conf/issta/GaleottiRPF10}. Assertion inference is the problem of generating an assertion from existing software artifacts, e.g., documentation, source code, etc. It is closely related to the oracle problem~\cite{DBLP:journals/tse/BarrHMSY15}, i.e., deciding whether or not a program execution is coherent with the desired behavior of the program. Assertion inference has many applications in software development including testing, e.g., to verify the expected outcome of a given test case~\cite{DBLP:journals/tse/FraserZ12}. Assertions can also capture program properties that should hold at specific program locations. In this paper, we focus on \emph{postcondition} based assertions that define the expected properties that must hold at the end of a given function's execution.

Figure~\ref{fig_assertion_inference} depicts the process that existing assertion inference techniques~(\cite{DBLP:conf/sigsoft/TerragniJTP20, DBLP:conf/icse/MolinaPAF21, DBLP:conf/icse/MolinadA22}) follow to infer assertions. First, based on an assertion generation approach utilized (e.g. GAssert~\cite{DBLP:conf/sigsoft/TerragniJTP20} and EvoSpex~\cite{DBLP:conf/icse/MolinaPAF21} use evolutionary search algorithm, SpecFuzzer~\cite{DBLP:conf/icse/MolinadA22} uses fuzzing), a technique generates candidate assertions for a given program/function. Then, the program's test suite is executed to determine which of those assertions are consistent with the behaviours exhibited by the actual program behaviour. Lastly, the validated assertions (i.e., those that are consistent with the test suite executions) go through the mutation analysis process for filtration of weak assertions. Here, a validated assertion that is also consistent with all the mutants' execution of a given program, is considered as weak because it is unable to distinguish between  the correct and a buggy program behaviour, and is hence discarded. The inferred assertions are the ones that are coherent with the actual program behaviour but do not satisfy the buggy program behaviour (at least kill 1 mutant). In the following section, we elaborate further on the existing techniques.

\subsection{Assertion Inference Techniques}
\label{subsec_assertion_inference_techniques}

\emph{Daikon}~\cite{DBLP:journals/scp/ErnstPGMPTX07} is a dynamic technique that infers assertions by monitoring test executions. Given a program under analysis, Daikon requires a test suite in order to infer specifications for such program. It uses the test suite to exercise the program, monitors program states at various program points, and then considers a set of candidate assertions obtained by instantiating assertion patterns. Those assertions that are \emph{not invalidated} by any test at a given program point are reported as likely invariants at the program point.

\emph{GAssert}~\cite{DBLP:conf/sigsoft/TerragniJTP20} and \emph{EvoSpex}~\cite{DBLP:conf/icse/MolinaPAF21} are assertion inference techniques based on evolutionary search algorithms. Similar to Daikon, these tools execute a test suite of the program under analysis and observe the  execution to infer assertions that are consistent with the observations. 
The components of their evolutionary processes are specifically designed to handle their respective assertion language supported, and thus, changing or extending the assertion languages implies redefining the corresponding evolutionary operators and other elements of the evolutionary processes, which is a non-trivial task.

\emph{SpecFuzzer}~\cite{DBLP:conf/icse/MolinadA22} is a recently proposed assertion inference technique that \emph{outperforms} the previous techniques. It uses a combination of static analysis, grammar-based fuzzing, and mutation analysis to infer assertions. First, it uses a lightweight static analysis to produce a grammar for the assertion language, which is tuned to the software under analysis. Second, it uses a grammar-based fuzzer to generate candidate assertions from that grammar. Then, a dynamic detector determines which of those assertions are consistent with the behavior exhibited by a provided test suite. In the final step, which is consistent with the previous techniques, SpecFuzzer eliminates redundant and irrelevant assertions using a selection mechanism based on mutation analysis. A salient feature of SpecFuzzer is that developers can easily adjust the specifications produced by tuning the grammar as opposed to making changes in the tool.

\subsection{\aim}
\label{subsec_aim}
\emph{Candidate assertions} that assertion inference techniques generate undergo a two-step filtering process (see Figure~\ref{fig_assertion_inference}). In the first step, the test suite of a target class \emph{C} falsifies the assertions that are invalid, i.e. are not satisfied by the legit program behaviour that the test suite executes. Though important to identify \emph{valid assertions}, such  filtering is not enough as it leaves room for \emph{weak assertions}, i.e, assertions that are trivial to satisfy and would not trigger any error if the target class \emph{C} had incorrect behaviour. For instance, a tautology, such as \begin{math}assert(x >= y~||~x <= y)\end{math},  it is a valid proposition that cannot be falsified, but it is unlikely to be useful. In the case of SpecFuzzer~\cite{DBLP:conf/icse/MolinadA22}, the fuzzer reports thousands of constraints, (i.e., candidate assertions), and only a few are invalidated by the test suite. Such weak assertions are not useful and should be discarded.

Mutation analysis is used to discard weak assertions~\cite{DBLP:conf/sigsoft/TerragniJTP20, DBLP:conf/icse/MolinaPAF21, DBLP:conf/icse/MolinadA22}. In general, the underlying idea is that valid assertions that are also coherent with every mutant's execution of target class \emph{C} are weak because they represent properties that hold also for buggy versions of \emph{C} (the mutants). On the contrary, assertions that do not hold for at least one mutant of \emph{C}, are useful because they are capable of distinguishing buggy versions of the code, aka mutants. We refer to such mutants that are killed by candidate assertions as \aim.


Despite its effectiveness in discarding weak assertions, mutation analysis suffers from scalability issues because many mutants can be generated from even a small piece of code, and most of these mutants are redundant. This adversely affects the overall performance of assertion inference techniques, especially on large subjects. To deal with this problem, we introduce \our, a \emph{static} technique that predicts \aim without requiring any dynamic analysis and aims to enhance efficiency of assertion inference techniques.

\subsection{Subsuming Mutants}
\label{subsec_subsuming_mutants}
Mutation analysis is computationally expensive even beyond its use for assertion inference. This is mainly due to the large number of mutants that it introduces, all of which require analysis and execution. In traditional mutation testing -- where the goal is to assess the ability of a test suite to ``kill'' mutants (i.e., to distinguish the observable behavior between the mutant and the original program) -- one can reduce the number of mutants to analyze by identifying the \emph{subsuming mutants}~\cite{DBLP:conf/apsec/KintisPM10, DBLP:conf/icst/AmmannDO14, 9677967}. Given two mutants $M_1$ and $M_2$, $M_1$ subsumes $M_2$ if every test case $T$ killing $M_1$ also kills $M_2$. Then, the computational cost of mutation analysis can be reduced by identifying the minimal subset of subsuming mutants, such that any test suite able to kill these mutants can also kill the entire set of killable mutants (excluding mutants that are functionally equivalent to the original program and cannot be killed). Hence, practitioners can perform mutation testing efficiently by analyzing only subsuming mutants. 

Given the potential of subsuming mutants in reducing mutation testing overheads, we investigate whether they are suitable for assertion inference (can help filter weak assertions). As we discuss in Section~\ref{subsec_rq1}, subsuming mutants are not sufficient for the assertion inference task as their use results in losing almost half of the inferred assertions (compared to considering all mutants).

\begin{figure}[t]
\centerline{\includegraphics[width=0.4\textwidth]{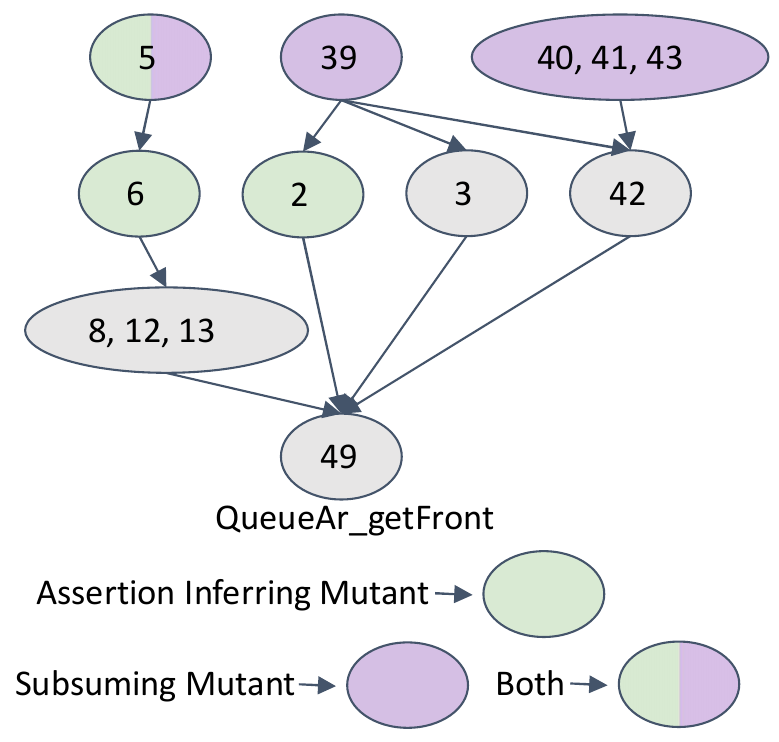}}
\caption{Mutant subsumption hierarchy for subject QueueAr\_getFront showing the positions of \aim and \emph{Subsuming Mutants}}
\label{fig_motivation}
\end{figure}

\section{Illustrative Example}
\label{sec_motivation}

Figure~\ref{fig_motivation} shows the mutants generated for the function \emph{getFront()} of class \emph{QueueAr}, one of our test subjects. The graph depicts the mutants' subsumption hierarchy, which is a standard way to represent subsumption relations between a set of mutants generated for the same code. Here, nodes represent mutants of the function and every edge connects mutants to other mutants that the former subsume. In our example, mutant 39 subsumes mutants 2, 3 and 42. Mutually subsuming mutants are also represented in the same node -- e.g. 40, 41 and 43. In this figure, we highlight in purple which mutants are subsuming mutants (at the top of the hierarchy) and in green which ones are \aim.

We execute SpecFuzzer~\cite{DBLP:conf/icse/MolinadA22} to infer assertions for subject \emph{QueueAr\_getFront} with its default configuration, i.e., by using all mutants available. 
SpecFuzzer infers \emph{27} assertions while the assertion filtering step via mutation analysis (rightmost part of Figure~\ref{fig_assertion_inference}) took 91 minutes on our infrastructure (see Section~\ref{sec_experimental_setup}). By contrast, using only subsuming mutants in the filtering step takes only 2.5 minutes (36.4 times faster) but would only produce 5 assertions.

These results confirm that, while reducing the number of mutants to analyze can improve the computational efficiency of the filtering process, subsuming mutants are not appropriate for this task. Intuitively, this is because the initial purpose of subsuming mutants is to minimize the number of tests needed to kill all mutants. In the context of assertion inference one rather aims to infer all valid assertions that can distinguish the mutants from the original code, that is, generate as many assertions that capture the specific code properties. For instance, in our \emph{QueueAr\_getFront} example, mutant \emph{5} satisfies all valid assertions except for five of them. In other words, considering mutant \emph{5} for analysis would result in inference of only 5 assertions. On the other hand, mutant \emph{6} filters 21 valid assertions, which means that considering mutant \emph{6} for analysis would result in inference of 21 assertions. Considering only subsuming mutants for analysis discards mutant \emph{6} as it is subsumed by mutant \emph{5} and hence it results in losing 21 strong assertions that could have been inferred.

The above example demonstrates the difference between \emph{Subsuming Mutants} and \aim, and the need for an approach that can efficiently identify the latter in order to save time on the mutation analysis step while maintaining the benefits of assertion inference. We propose \our, the first mutant selection method for assertion inference. Applying it with SpecFuzzer on the \emph{QueueAr\_getFront} example, \our predicts mutant \emph{6} as assertion inferring mutant and helps to infer 21 assertions (out of 27 assertions when using all mutants), for only a fraction of the computation time, i.e., 30 seconds (instead of 91 minutes taken to analyze all mutants).

\begin{figure*}[ht]
\centerline{\includegraphics[width=0.98\textwidth]{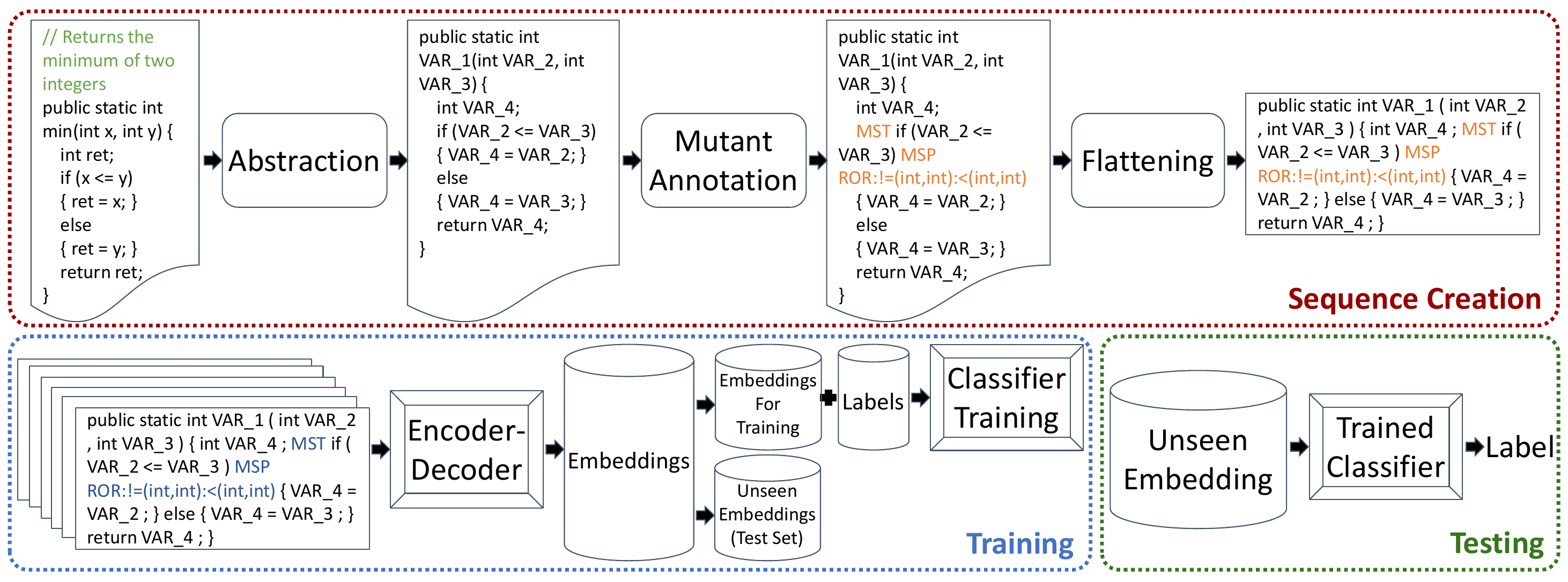}}
\caption{Overview of \our: Source code is abstracted and annotated to represent a mutant, which is further flattened to create a space separated sequence of tokens. An encoder-decoder model is trained on token sequences to generate mutant embeddings. A classifier is trained on these embeddings and their corresponding labels (whether or not the mutant is assertion inferring). The trained classifier can then be used for label prediction of an unseen mutant.}
\label{fig_implementation}
\end{figure*}

\section{Approach}
\label{sec_approach}

The main objective of \our is to predict whether a mutant (of a previously unseen piece of code) is likely to be assertion inferring. In order for our approach to be lightweight in terms of engineering and computational effort, we want \our to be able to (a) learn relevant features of \aim without requiring manual feature definition, and (b) do so without costly dynamic analysis of mutant executions. To achieve this, we decompose our problem into two parts: learn a representation of mutants using code embedding techniques, and learn to predict, based on such embeddings, whether the represented mutants are \aim. 

\subsection{Overview of \our}

Figure~\ref{fig_implementation} shows an overview of \our. We decompose our approach into three steps that we detail later in this section:
\begin{enumerate}
\item \emph{Build a token representation}: \our pre-processes the original code in order to remove irrelevant information and produce abstracted code, which is then tokenized to form a sequence of tokens. Each mutant is ultimately transformed into its corresponding token representation and undergoes the next step.  
\item \emph{Representation learning}: We train an encoder-decoder model to generate an embedding, aka vector representation of the mutant. This step is where \our automatically learns the relevant features of mutants without requiring an explicit definition of these features. 
\item \emph{Classification:} \our trains a classification model to classify the mutants (based on their embeddings) as \aim or not. The true labels used for training the model are obtained by running SpecFuzzer on the original code, and checking whether the mutants are \aim with respect to the candidate (and test-suite validated) assertions that SpecFuzzer generates. 
\end{enumerate}

It is interesting to note that the mutant representation learned by \our does not depend on the particular set of assertions that SpecFuzzer (or any other assertion inference technique) would check against the mutant. \our rather aims to learn properties of the mutants (and their surrounding context) that are generally useful for assertion inference. This is in line with the recent work on contextual mutant selection \cite{JustKA17, 9677967, ChekamPBTS20} that aims at selecting high utility mutants for mutation testing. This characteristics makes \our applicable to pieces of code that it has not seen during training. In particular, our experiments reveal the capability of \our to be effective on projects not seen during training. Certainly, the assertion inference technique that we use to build the true labels in the classification tasks is important because this technique should produce a sufficiently large set of useful assertions -- an essential condition for our classifier to provide relevant prediction results. We use SpecFuzzer~\cite{DBLP:conf/icse/MolinadA22} for its state of the art performance, i.e., SpecFuzzer outperforms the existing techniques (GAssert~\cite{DBLP:conf/sigsoft/TerragniJTP20} and EvoSpex~\cite{DBLP:conf/icse/MolinaPAF21}) in assertion inference (SpecFuzzer infers 7 times and 15 times more assertions than GAssert and EvoSpex) and achieves better performance with respect to the ground truth by achieving better Recall and F-1 score than the existing.

\subsection{Token Representation}

A major challenge in learning from raw source code is the huge vocabulary created by the abundance of identifiers and literals used in the code \cite{DBLP:journals/tosem/TufanoWBPWP19, DBLP:conf/icse/TufanoPWBP19, DBLP:conf/iclr/AlonBLY19}. In our case, this large vocabulary may hinder \our's ability to learn relevant features of \aim. Thus, we first abstract original (non-mutated) source code by replacing user-defined entities (function names, variable names, and string literals) with generic identifiers that can be reused across the source code file. During this step, we also remove code comments. This pre-processing yields an abstracted version of the original source code, as the abstracted code snippet in Figure~\ref{fig_implementation}.

To perform the abstraction, we use the publicly available tool \emph{src2abs}~\cite{DBLP:conf/icse/TufanoPWBP19}. This tool first discerns the type of each identifier and literal in the source code. Then, it replaces each identifier and literal in the stream of tokens with a unique ID representing the type and role of the identifier/literal in the code. Each ID \textless\texttt{TYPE}\textgreater\texttt{$\_\#$} is formed by a prefix, (i.e., \textless\texttt{TYPE}\textgreater\texttt{$\_$} ) which represents the type and role of the identifier/literal, and a numerical ID, (i.e.,~\texttt{$\#$}) which is assigned sequentially when reading the code. These IDs are reused when the same identifier/literal appears again in the stream of tokens. Although we use src2abs, as an alternative, one can use any utility that identifies user-defined entities and replaces such with reusable identifiers.

Next, to represent a mutant, we annotate the abstracted code with a mutation annotation on the statement where the mutation is to be applied. These annotations have the general shape ``\texttt{MST}~statement~\texttt{MSP~MutationOperator}'', where \texttt{MST} and \texttt{MSP} denote mutation annotation start and stop, respectively, and these are followed by a \texttt{MutationOperator} that indicates the applied mutation operation (as shown in figure~\ref{fig_implementation}). We repeat the process for every mutant.

Finally, we flatten every mutant (by removing newline, extra whitespace, and tab characters) to create a single space separated sequence of tokens. Using these sequences, we intend to capture as much code as possible around the mutant without incurring an exponential increase in training time~\cite{DBLP:conf/icsm/TufanoWBPWP19, DBLP:conf/icse/TufanoPWBP19, 9677967, DBLP:journals/ese/GargDJCPT22}, we found a sequence length of 500 tokens to be a good fit for our task as it does not exceed 24 hours of training time (wall clock) on a Tesla V100 GPU.


\subsection{Embedding Learning with Encoder-Decoder}

Our next step is to learn embedding, aka vector representation (that can later be used to train a classification model) from mutants' token representation. We develop an encoder-decoder model, a neural architecture commonly used in representation learning task~\cite{DBLP:conf/emnlp/KalchbrennerB13}. The key principles of our encoder-decoder architecture is that the encoder transforms the token representation into an embedding and the decoder attempts to retrieve the original token representation from the encoded embedding. The learning objective is then to minimize the binary cross-entropy between the original token representation and the decoded one. Once the model training has converged, we can compute the embedding from any other mutant's token representation by feeding the latter into the encoder and retrieve the output.

We use a bi-directional Recurrent Neural Network (RNNs)~\cite{DBLP:journals/corr/BritzGLL17} to develop our encoder-decoder, as previous works on code learning have demonstrated the effectiveness of these models to learn useful representations from code sequences~\cite{DBLP:journals/corr/BahdanauCB14, 9677967, DBLP:journals/ese/GargDJCPT22, DBLP:conf/nips/SutskeverVL14}. We build \our on top of \emph{tf-seq2seq}~\cite{DBLP:journals/corr/AbadiABBCCCDDDG16}, an established general-purpose encoder-decoder framework. We use a Gated Recurrent Units (GRU) network~\cite{DBLP:conf/emnlp/ChoMGBBSB14} to act as the RNN cell, which was shown to perform better than simpler alternatives (e.g. simple RNNs) both in software engineering and other learning tasks~\cite{DBLP:journals/jaiscr/ShewalkarNL19, 9677967}. To achieve good performance with acceptable model training time, we utilize AttentionLayerBahdanau \cite{DBLP:conf/icassp/BahdanauCSBB16} as our attention class, configured with 2 layered AttentionDecoder and 1 layered BidirectionalRNNEncoder, both with 256 units. 

To determine an appropriate number of training epochs for model convergence, we conducted a preliminary study involving a small validation set (independent of both the training and test sets used in our evaluation) where we monitor model's performance in replicating (as output) the same mutant sequence provided as input. We pursue training the model till the training performance on the validation set does not improve anymore. We found 10 epochs for the sequences up to a length of 500 tokens to be a good default for our validation sets.

\subsection{Classifying \aim}

Next, we train a classification model in predicting whether a mutant (represented through the embedding produced by the RNN encoder) is likely to be \aim. The learning objective here is to maximize the classification performance (which we mainly measure with Matthews Correlation Coefficient (MCC), see Section \ref{subsec_performance_measurement}). To obtain our true classification labels, we run an assertion inferring technique (viz. SpecFuzzer) using all available mutants and exhaustively determine which mutants are assertion inferring. As for the classification model, we rely on \emph{random forests}~\cite{DBLP:journals/ml/Breiman01} because these are lightweight to train and have shown to be effective in solving various software engineering tasks~\cite{Jimenez2019realworld,Pinto2020vocabulary}. We used standard parameters for random forests, viz. we set the number of trees to 100, use Gini impurity for splitting, and set the number of features (i.e. embedding logits) to consider at each split to the square root of the total number of features.

Once the model training has converged, we can use the random forest to predict whether an unseen mutant is likely to be \aim. We make the mutant go through the pre-processing pipeline to obtain its abstract token representation, then feed it into the encoder-decoder architecture to retrieve its embedding and finally input it into the classifier to obtain the predicted label (\aim or not).


\section{Research Questions}
\label{sec_research_questions}
We start our analysis by investigating whether \aim can be approximated by other sets of mutants, such as \emph{randomly selected} and \emph{Subsuming mutants}, and contrast their performance with \our in the context of assertion inference. We compare with random mutant selection since it is an untargeted method that is often superior to many mutant selection strategies \cite{GopinathAAJG16, ZhangHHXM10} and is considered by the literature as a strong baseline \cite{KurtzAODKG16, 9677967, ChekamPBTS20}. We also compare with subsuming mutants since they form the main objective of mutant selection~\cite{KurtzAODKG16,9677967,PapadakisHHJT16} with numerous strategies targeting them~\cite{MarcozziBKPPC18, JustKA17, GongZYM17, 9677967}. Hence, we check the effectiveness (completeness w.r.t. to using all mutants) and efficiency (how much time is required) of SpecFuzzer~\cite{DBLP:conf/icse/MolinadA22}, a state of the art assertion inference technique, when utilizing mutant subsets over all supported mutants. Therefore we ask:
\begin{enumerate}
\item [\textbf{RQ1}] \emph{Performance Evaluation:} How effective and efficient is \our in comparison to subsuming, randomly selected and all mutants baseline methods with respect to the assertion inference task? 
\end{enumerate}
For this task, we considered the dataset provided by \emph{Molina et al.}~\cite{DBLP:conf/icse/MolinadA22}. We re-executed SpecFuzzer on 40 subjects, initially without discarding any mutant, and later by selecting the mutants following \our and our two baseline mutant selection techniques (subsuming and random mutant selection). In their work~\cite{DBLP:conf/icse/MolinadA22}, \emph{Molina et al.} carefully studied the subjects and manually produced corresponding (complete) \emph{Ground Truth} assertions capturing the intended behavior of the subjects. In our execution of SpecFuzzer, it was able to infer the ground truth assertions for 26 subjects, when all mutants were considered for assertion inference. Hence, we also compared the effectiveness of all three mutant selection techniques (as explained in the RQ1) in inferring \emph{Ground Truth} assertions. Hence, we ask:
\begin{enumerate}
\item [\textbf{RQ2}] \emph{Ground Truth Evaluation:} How \our compares with the subsuming and randomly selected mutants in terms of inferred ground truth assertions?
\end{enumerate}
In the above questions, comparisons between the three mutant selection techniques were feasible because SpecFuzzer inferred assertions (at least one) when considering all mutants. Now, we investigate if \our's predicted \aim can help SpecFuzzer to scale, i.e., if SpecFuzzer can infer assertions by considering only \our's predicted mutants in scenarios where SpecFuzzer timed out during mutation analysis and was not able to infer any assertion when all mutants were considered for analysis. For this task, we conducted experiments on 6 subjects from GitHub (table~\ref{tab_data}) where SpecFuzzer timed out. We also compared SpecFuzzer's performance when it considered \our's predicted mutants vs an equal number of randomly selected mutants (state of the art in mutant selection). Hence, we ask:
\begin{enumerate}
\item [\textbf{RQ3}] \emph{Scalability Evaluation:} Can \our improve the scalability of assertion inference techniques?
\end{enumerate}
\begin{table*}[t]
\caption{The table records the test subjects, Method details, All Mutants count, Assertion Inferring Mutants count, All Assertions and Ground Truth Assertions inferred when all mutants are used, (i.e., Specfuzzer's default execution with no mutant selection)}
\begin{center}
\begin{tabular}{l|l|c|c|c|c}
\toprule
\multicolumn{1}{c|}{\textbf{Subject}} & \multicolumn{1}{|c|}{\textbf{Method}} & \textbf{All} & \textbf{Assertion} & \textbf{All} & \textbf{Ground} \\
 &  & \textbf{Mutants$\#$} & \textbf{Inferring} & \textbf{Assertions$\#$} & \textbf{Truth} \\
 &  &  & \textbf{Mutants$\#$} &  & \textbf{Assertions$\#$} \\ \midrule
ArithmeticUtils$\_$subAndCheck & math.ArithmeticsUtils.subAndCheck & 16 & 2 & 3 & 1 \\ 
BooleanUtils$\_$compare & lang.BooleanUtils.compare & 13 & 13 & 29 & 3 \\ 
composite$\_$addChild & eiffel.Composte.addChild & 35 & 6 & 185 & 0 \\ 
doublylinkedlistnode$\_$insertRight & eiffel.DLLN.insert$\_$right & 18 & 7 & 16 & 2 \\ 
doublylinkedlistnode$\_$remove & eiffel.DLLN.remove & 18 & 4 & 21 & 1 \\ 
Envelope$\_$maxExtent & tsuite.Envelope.maxExtent & 56 & 10 & 188 & 0 \\ 
FastMathNew$\_$floor & math.FastMath.floor & 42 & 18 & 60 & 2 \\ 
IntMath$\_$mod & guava.IntMath.mod & 21 & 15 & 199 & 0 \\ 
listcomp02$\_$insert$\_$r & cozy.ListComp02.insert$\_$r & 20 & 2 & 1 & 0 \\ 
listcomp02$\_$insert$\_$s & cozy.ListComp02.insert$\_$s & 20 & 1 & 1 & 0 \\ 
map$\_$count & eiffel.Map.count & 63 & 3 & 4 & 0 \\ 
map$\_$extend & eiffel.Map.extend & 65 & 9 & 10 & 3 \\ 
map$\_$remove & eiffel.Map.remove & 63 & 1 & 1 & 0 \\ 
MathUtilsNew$\_$copySignInt & math.MathUtils.copySignInt & 48 & 2 & 16 & 0 \\ 
MathUtil$\_$clamp & tsuite.MathUtil.clamp & 11 & 8 & 12 & 3 \\ 
maxbag$\_$add & cozy.MaxBag.add & 748 & 53 & 49 & 1 \\ 
maxbag$\_$getMax & cozy.MaxBag.get$\_$max & 749 & 21 & 25 & 1 \\ 
maxbag$\_$remove & cozy.MaxBag.remove & 748 & 67 & 26 & 1 \\ 
polyupdate$\_$a1 & cozy.PolyUpdate.a & 54 & 26 & 100 & 2 \\ 
polyupdate$\_$sm & cozy.PolyUpdate.sm & 56 & 13 & 73 & 1 \\ 
QueueAr$\_$dequeue & daikon.QueueAr.dequeue & 66 & 9 & 68 & 3 \\ 
QueueAr$\_$dequeueAll & daikon.QueueAr.dequeueAll & 67 & 11 & 69 & 1 \\ 
QueueAr$\_$enqueue & daikon.QueueAr.enqueue & 66 & 17 & 119 & 2 \\ 
QueueAr$\_$getFront & daikon.QueueAr.getFront & 67 & 3 & 27 & 0 \\ 
QueueAr$\_$makeEmpty & daikon.QueueAr.makeEmpty & 67 & 20 & 73 & 1 \\ 
ringbuffer$\_$count & eiffel.RingBuffer.count & 101 & 28 & 119 & 0 \\ 
ringbuffer$\_$extend & eiffel.RingBuffer.extend & 101 & 20 & 148 & 0 \\ 
ringbuffer$\_$item & eiffel.RingBuffer.item & 101 & 11 & 116 & 0 \\ 
ringbuffer$\_$remove & eiffel.RingBuffer.remove & 101 & 14 & 143 & 0 \\ 
ringbuffer$\_$wipeOut & eiffel.RingBuffer.wipe$\_$out & 101 & 13 & 95 & 1 \\ 
simple-examples$\_$abs & oasis.SimpleMethods.abs & 20 & 18 & 30 & 1 \\ 
simple-examples$\_$addElementToSet & oasis.SimpleMethods.addElementToSet & 3 & 2 & 1 & 1 \\ 
simple-examples$\_$getMin & oasis.SimpleMethods.getMin & 7 & 6 & 51 & 1 \\ 
StackAr$\_$makeEmpty & daikon.StackAr.makeEmpty & 47 & 13 & 47 & 1 \\ 
StackAr$\_$pop & daikon.StackAr.pop & 63 & 10 & 35 & 2 \\ 
StackAr$\_$push & daikon.StackAr.push & 55 & 6 & 25 & 2 \\ 
StackAr$\_$top & daikon.StackAr.top & 50 & 8 & 3 & 0 \\ 
StackAr$\_$topAndPop & daikon.StackAr.topAndPop & 54 & 13 & 68 & 2 \\ 
structure$\_$foo & cozy.Structure.foo & 27 & 5 & 1 & 1 \\ 
structure$\_$setX & cozy.Structure.setX & 26 & 15 & 131 & 1 \\ 
EmailScanner$\_$findFirst & org.nibor.autolink.internal.EmailScanner.findFirst & 134 & \multicolumn{3}{|c}{Scalability Evaluation (RQ3)*} \\ 
EmailScanner$\_$scan & org.nibor.autolink.internal.EmailScanner.scan & 134 & \multicolumn{3}{|c}{Scalability Evaluation (RQ3)*} \\ 
IdentityHashSet$\_$isEmpty & org.leplus.ristretto.util.IdentityHashSet.isEmpty & 23 & \multicolumn{3}{|c}{Scalability Evaluation (RQ3)*} \\ 
OptionGroup$\_$setRequired & org.apache.commons.cli.OptionGroup.setRequired & 34 & \multicolumn{3}{|c}{Scalability Evaluation (RQ3)*} \\ 
OptionGroup$\_$setSelected & org.apache.commons.cli.OptionGroup.setSelected & 34 & \multicolumn{3}{|c}{Scalability Evaluation (RQ3)*} \\ 
Scanners$\_$findUrlEnd & org.nibor.autolink.internal.Scanners.findUrlEnd & 111 & \multicolumn{3}{|c}{Scalability Evaluation (RQ3)*} \\ 
\bottomrule
\multicolumn{6}{l}{* Subjects for which SpecFuzzer timed out during mutation analysis are considered for Scalability Evaluation (RQ3).}
\end{tabular}
\label{tab_data}
\end{center}
\end{table*}

\section{Experimental Setup}
\label{sec_experimental_setup}

\subsection{Data and Tools}
We selected 46 Java methods; 40 subjects that were used in previous studies~\cite{DBLP:conf/sigsoft/TerragniJTP20, DBLP:conf/icse/MolinaPAF21, DBLP:conf/icse/MolinadA22} for evaluating performance in RQ1 and RQ2, and 6 larger subjects from GitHub for the scalability evaluation in RQ3. In their study, \emph{Molina et al.}~\cite{DBLP:conf/icse/MolinadA22} manually constructed \emph{Ground Truth} assertions capturing the intended behavior of these 40 subjects. We use these assertions to answer RQ2. 
Table~\ref{tab_data} records the details of our dataset. 

To perform mutation testing we used Major~\cite{DBLP:conf/kbse/JustSK11} mutation testing tool and to construct comprehensive test suites (and improve the chances to infer true assertions) we used EvoSuite~\cite{DBLP:conf/sigsoft/FraserA11} and Randoop~\cite{DBLP:conf/icse/PachecoLEB07} to augment the developer test suites, similarly to what was done by previous work~\cite{DBLP:conf/icse/MolinadA22}.  


\subsection{Prediction Performance Metrics}
\label{subsec_performance_measurement}
\aim prediction modeling is a binary classification problem, thus it can result in four types of outputs: Given a mutant is assertion inferring, if it is predicted as assertion inferring, then it is a true positive (TP); otherwise, it is a false negative (FN). Vice-versa, if a mutant does not infer any assertion and, if it is predicted as assertion inferring then it is a false positive (FP); otherwise, it is a true negative (TN). From these, we can compute the traditional evaluation metrics such as \emph{Precision} and \emph{Recall}, which quantitatively evaluate the prediction accuracy of prediction models.
\begin{align*}
\emph{Precision} = \frac{TP}{TP + FP}  \hspace{1em} \emph{Recall} = \frac{TP}{TP + FN} 
\end{align*}
Intuitively, \emph{Precision} indicates the ratio of correctly predicted positives over all the considered positives. \emph{Recall} indicates the ratio of correctly predicted positives over all actual positives. 
Yet, these metrics do not take into account the true negatives and can be misleading, especially in the case of imbalanced data. Hence, we complement these with the \emph{Matthews Correlation Coefficient (MCC)}, a reliable metric of the quality of prediction models~\cite{DBLP:conf/ease/YaoS20}. It is regarded as a balanced measure that can be used even when the classes are of very different sizes~\cite{DBLP:journals/tse/ShepperdBH14}, e.g. \emph{12.95\%} \aim in total, for 40 subjects in our dataset (table~\ref{tab_data}). \emph{MCC} is calculated as: \[\emph{MCC} = \frac{TP \times TN - FP \times FN}{\sqrt{(TP + FP)(TP + FN)(TN + FP)(TN + FN)}}\] 
\emph{MCC} returns a coefficient between 1 and -1. An MCC value of 1 indicates a perfect prediction, while a value of -1 indicates a perfect inverse prediction, i.e., a total disagreement between prediction and reality. MCC value of 0 indicates that the prediction performance is equivalent to random guessing.

\subsection{Experimental Procedure}
\label{subsec:experimentalsettings}
To answer our RQs we executed SpecFuzzer to infer assertions for all subjects (Table~\ref{tab_data}) with its default configuration, i.e., using all mutants to filter candidate assertions during the mutation testing step (Figure~\ref{fig_assertion_inference}). 
We also determined \aim and \emph{Subsuming Mutants} from SpecFuzzer execution logs for the 40 subjects used in RQ1 and RQ2. 
Once we labeled, we re-execute SpecFuzzer by employing the following 3 mutant selection techniques: 

\begin{itemize}
    \item \sms. We execute SpecFuzzer by only considering subsuming mutants for mutation analysis and discarding the rest of the mutants from the original set. 

    \item \our. We train models on \aim and perform k-fold cross validation (where k = 5) at the project level, i.e., we train on 32 subjects and evaluate/test on 8 unseen during testing subjects. Once we get the predictions for all 40 subjects, we re-execute SpecFuzzer by only considering the predicted mutants and by discarding all other mutants from the original set. 

    \item \rnd. We randomly select an equal number of mutants (equal to the number of predicted mutants) from the original set of mutants and re-execute SpecFuzzer by only considering these randomly selected and by discarding all other mutants. We repeat this step 10 times to eliminate the chances to report coincidental results. We report the median case results.  

\end{itemize}

To answer \emph{RQ1}, we compute the Prediction Performance Metrics of \our in order to show its learning ability. This is a sanity check that our prediction modeling framework indeed manages to predict something well. However prediction results does not reflect the end-task (assertion inference) performance since mutants are not independent, i.e., there are large overlaps between the tests and assertions that lead to mutant kills. This means that subsuming or randomly selected mutants may perform similarly to \our. We thus, measure the cost of the employed mutant selection technique, i.e., how many assertions are \emph{not} inferred (which are inferred when all mutants are considered), and the benefit gained, i.e., the improvement in assertion inference in terms of wall clock time. 

To answer \emph{RQ2}, we check the results of RQ1 and compare how many Ground Truth assertions SpecFuzzer infers with each mutant selection technique (Completeness).  It should be noted that it was able to infer the ground truth assertions for 26 subjects out of the 40, when all mutants were considered for mutation analysis. Hence, we analyze the results only for these 26 subjects.

To answer \emph{RQ3}, i.e., if \our's predicted \aim can help SpecFuzzer to infer assertions for 6 subjects where it was not able to infer any assertion (timed out when all mutants were considered for analysis), we retrain \our on all 40 subjects (with available labeled mutants) and predict likely \aim for these 6 subjects. We re-execute SpecFuzzer by only using the predicted mutants and by discarding all other mutants from the original set. Additionally, we randomly select mutants in a similar fashion as before (following RQ1 experimental procedure) and re-execute SpecFuzzer accordingly to compare performance with \rnd. Thus to answer \emph{RQ3} we measure 1) in how many subjects, the selected mutants lead to assertion inference, and 2) the ratio of assertion inferring mutants from the entire set of mutants.

\section{Experimental Results}
\label{sec_experimental_results}
\begin{table}[t]
\caption{RQ1 results - Performance of Assertion Inference}
\begin{center}
\resizebox{.5\textwidth}{!}{
\begin{tabular}{l|c|c|c}
\multicolumn{4}{c}{\textbf{Mutation filtered assertion inference}} \\ 
\toprule
 & \textbf{With Subsuming} & \textbf{With} & \textbf{With Random} \\
 & \textbf{Mutant Selection} & \textbf{\our} & \textbf{Mutant Selection} \\ \midrule
\textbf{Inferred Assertions} & 57.77\% & 87.51\% & 51.47\% \\
(per Subject) &  &  &  \\ 
\textbf{Missed Assertions} & 42.23\% & 12.49\% & 48.53\% \\
\textbf{(Cost)} &  &  &  \\ 
\textbf{Improvement in} & 19.16 times & 46.29 times & 47.34 times \\
\textbf{Time (Benefit)} &  &  &  \\ \midrule
\multicolumn{4}{c}{} \\
\multicolumn{4}{c}{\textbf{Subjects with assertions inferred}} \\ \midrule
 Total Subjects\# 40 & \textbf{With Subsuming} & \textbf{With} & \textbf{With Random} \\
 & \textbf{Mutant Selection} & \textbf{\our} & \textbf{Mutant Selection} \\ \midrule
\textbf{Subjects with All} & 5 & 23 & 2 \\ 
\textbf{assertions inferred} &  &  &  \\ 
\textbf{Subjects with No} & 7 & 0 & 2 \\ 
\textbf{assertion inferred} &  &  &  \\ \bottomrule
\end{tabular}
}
\label{tab_rq1}
\end{center}
\end{table}

\begin{figure}[t]
\centerline{\includegraphics[width=0.45\textwidth]{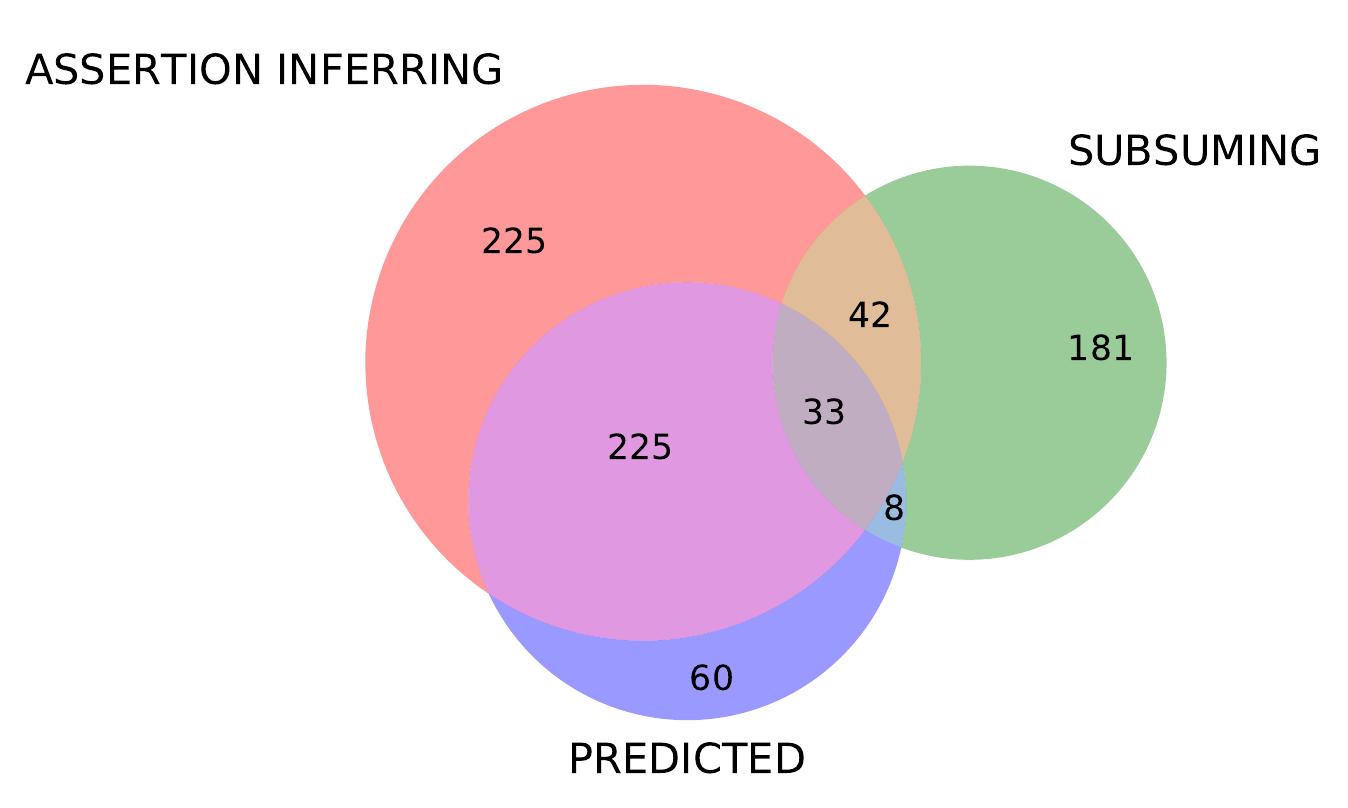}}
\caption{Mutant distribution} 
\label{fig_venn}
\end{figure}

\begin{table}[t!]
\caption{RQ2 Results - Inferring Ground Truth Assertions}
\begin{center}
\resizebox{0.5\textwidth}{!}{
\begin{tabular}{l|c|c|c}
\multicolumn{4}{c}{\textbf{Ground Truth assertion inference}} \\ \toprule
 & \textbf{With Subsuming} & \textbf{With} & \textbf{With Random} \\
 & \textbf{Mutant Selection} & \textbf{\our} & \textbf{Mutant Selection} \\ \midrule
\textbf{Inferred Assertions} & 67.31\% & 96.15\% & 19.23\% \\
(per Subject) &  &  &  \\ \midrule
\multicolumn{4}{c}{} \\
\multicolumn{4}{c}{\textbf{Subjects with assertions inferred}} \\ \midrule
 Total Subjects\# 26 & \textbf{With Subsuming} & \textbf{With} & \textbf{With Random} \\
 & \textbf{Mutant Selection} & \textbf{\our} & \textbf{Mutant Selection} \\ \midrule
\textbf{Subjects with All} & 17 & 25 & 5 \\ 
\textbf{assertions inferred} &  &  &  \\ 
\textbf{Subjects with No} & 8 & 1 & 21 \\ 
\textbf{assertion inferred} &  &  &  \\ \bottomrule
\end{tabular}
}
\label{tab_rq2}
\end{center}
\end{table}

\subsection{Performance Evaluation (RQ1)}
\label{subsec_rq1}

\our predicted \aim with a prediction performance of 0.58 MCC, 0.79 Precision, and 0.49 Recall. These values indicate that using \our should gain significant improvements in terms of inferred assertions over baseline methods. Figure~\ref{fig_venn} shows a Venn diagram recording the distribution of \aim, \our and \emph{Subsuming} mutant sets. The figure shows that a large number of \aim (450 out of 525) are not subsuming. At the same time \our detects almost half of them (258 out 525), indicating relatively good performance.

Table~\ref{tab_rq1} records SpecFuzzer's performance w.r.t. assertion inference by employing different mutant sets, i.e, \sms, \our, and \rnd. The results show that when SpecFuzzer uses \our's predicted mutants, it infers 87.51\% of total assertions, i.e. only 12.49\% missed assertions (the cost of considering only \our's predicted mutants) with 46.29 times faster mutation analysis (2.5 times faster than considering subsuming mutants). \our enables SpecFuzzer to infer at least one assertion for all subjects (inferring all assertions for 23 subjects). 

When SpecFuzzer uses the subsuming mutants, it infers 57.77\% of total assertions. It infers all assertions for 5 subjects but fails to infer any for 7 subjects. Although it misses 42.23\% of the assertions (the cost of considering only the subsuming mutants), it reaps the benefit of an improved mutant analysis time of 19.16 times faster than when using all mutants. 

A similar improvement, with subsuming mutants, in the mutation testing time is noted when SpecFuzzer uses randomly selected mutants, but it fails to infer 48.53\% of total assertions. In two cases it infers all assertions and fails to infer any assertion for 2 other cases. \our outperforms \rnd with a statistically significant\footnote{We compared the inferred assertion percentages using Wilcoxon sign-rank-test and obtained a $p-value < 5.39\mathrm{e}{-7}$ with \rnd.} sizeable difference.

\begin{tcolorbox}[standard jigsaw, opacityback=0]
Answer to RQ1: \our predicts \aim with 0.58 MCC value. \our enables SpecFuzzer to infer assertions for all subjects, running 46.29 times faster at the expense of 12.49\% of the assertions. At the same time, \our enables SpecFuzzer to infer 36\% and 30\% more assertions than \rnd and \sms, while runs 2.5 times faster than \sms and requires similar execution time (wall clock) to \rnd.
\end{tcolorbox}

\subsection{Completeness Evaluation (RQ2)}
\label{subsec_rq2}
Table~\ref{tab_rq2} records SpecFuzzer's performance in ground truth assertion inference by employing different mutant selection techniques, i.e, \sms, \our, and \rnd. When SpecFuzzer considers only subsuming mutants, it infers 67.31\% of all ground truth assertions (inferred without any mutant selection technique). It infers all ground truth assertions for 17 subjects but fails to infer any for 8 subjects. On considering \our's predicted mutants, SpecFuzzer infers almost all ground truth assertions, i.e, 96.15\%. \our's predicted mutants enable SpecFuzzer to infer at least one ground truth assertion for all subjects except for one subject. When SpecFuzzer considers randomly selected mutants, it infers 19.23\% of all ground truth assertions. It infers all assertions for 5 subjects whereas fails to infer assertion for 21 subjects. \our outperforms \rnd with a statistically significant\footnote{We compared the inferred ground truth assertion percentages and obtained a $p-value < 7.74\mathrm{e}{-6}$ with \rnd.} sizeable difference.

\begin{tcolorbox}[standard jigsaw, opacityback=0]
Answer to RQ2: \our's predicted mutants enable SpecFuzzer to infer ground truth assertions for almost all subjects except one, inferring 96.15\% of the total assertions which is superior to both \sms (infers 67.31\%) and \rnd  (infers 19.23\%).
\end{tcolorbox}

\begin{table}[t]
\caption{RQ3 results - Scalability Evaluation}
\begin{center}
\resizebox{.5\textwidth}{!}{
\begin{tabular}{l|c|c}
\multicolumn{3}{c}{\textbf{\aim (among mutants selected)}} \\ \toprule
Mutants selected: 2.99\% from the & \textbf{With} & \textbf{With Random} \\
entire mutant set (per subject) & \textbf{\our} & \textbf{Mutant Selection} \\ \midrule
\textbf{\aim} & 83.33\% & 16.67\% \\
(among selected mutants) &  &  \\ \midrule
\multicolumn{3}{c}{} \\
\multicolumn{3}{c}{\textbf{Inferred assertions\#}} \\ \midrule
 \textbf{Subject} & \textbf{With} & \textbf{With Random} \\
 & \textbf{\our} & \textbf{Mutant Selection} \\ \midrule
EmailScanner\_findFirst & 85 & 58 \\ 
EmailScanner\_scan & 192 & 0 \\ 
IdentityHashSet\_isEmpty & 3 & 2 \\ 
OptionGroup\_setRequired & 8 & 8 \\ 
OptionGroup\_setSelected & 8 & 0 \\ 
Scanners\_findUrlEnd & 23 & 0 \\ \bottomrule
\end{tabular}
}
\label{tab_rq3}
\end{center}
\end{table}

\subsection{Scalability Evaluation (RQ3)}
\label{subsec_rq3}
Table~\ref{tab_rq3} records the results of the SpecFuzzer's performance in inferring assertions when it employs  \our and \rnd, for the subjects where mutation testing timed out. \our selected 2.99\% mutants from the entire mutant set. Among the predicted mutants, 83.33\% mutants are assertion inferring. When an equal number of mutants are selected using \rnd, only 16.67\% of mutants selected are assertion inferring. When SpecFuzzer considers only \our's predicted mutants for assertion filtering, it infers assertions for all subjects as shown in table~\ref{tab_rq3} with a statistically significant\footnote{We compared the percentages of \aim among the selected mutants, using Wilcoxon sign-rank-test and obtained a $p-value < 9.98\mathrm{e}{-6}$ with \rnd.} sizeable difference. On the other hand, for 50\% of the subjects (3 out of 6), SpecFuzzer fails to infer any assertion if it uses \rnd.\\

\begin{tcolorbox}[standard jigsaw, opacityback=0]
Answer to RQ3: \our enables SpecFuzzer to scale by inferring assertions for all subjects where full mutation testing timed out and \rnd failed in  50\% of the cases.
\end{tcolorbox}
\section{Threats to Validity}
\label{sec_threats_to_validity}

\textit{External Validity}: Threats may relate to the subjects we used. Although our evaluation expands to projects of various sizes, the results may not generalize to other projects. We consider this threat of low importance since we have a large sample of subjects (40 subjects from the previous studies~\cite{DBLP:conf/sigsoft/TerragniJTP20, DBLP:conf/icse/MolinaPAF21, DBLP:conf/icse/MolinadA22} and 6 subjects from GitHub for scalability evaluation). Moreover, our predictions are based on the local mutant context, that has been shown to be determinant of mutants' utility \cite{JustKA17, 9677967}. Other threats may relate to the assertion inference technique that we used for evaluation. This choice was made since SpecFuzzer is the current state of the art and operates similarly to other techniques (the main differences lie in the grammar used). We consider this threat of low importance since \our deals with mutation testing, which is used in the same way by all assertion inference techniques~\cite{DBLP:conf/sigsoft/TerragniJTP20, DBLP:conf/icse/MolinaPAF21, DBLP:conf/icse/MolinadA22}, and are directly impacted by the number of mutants involved. Nevertheless, in case other techniques require different predictions, one could re-train, tune and use  \our for the specific method of interest, as we did here with SpecFuzzer. 

\textit{Internal Validity}: Threats may relate to the restriction that we impose on sequence length, i.e., a maximum of \emph{500} tokens. This was done to enable reasonable model training time, approximately \emph{24} hours to learn mutant embeddings on Tesla V100 gpu. Other threats may be due to the use of \emph{tf-seq2seq}~\cite{DBLP:journals/corr/AbadiABBCCCDDDG16} for learning mutant embeddings. This choice was made for simplicity, to use the related framework out of the box, similar to the related studies \cite{DBLP:conf/icse/TufanoPWBP19, DBLP:journals/ese/GargDJCPT22}. Other internal validity threats could be related to the test suites we used and the mutants considered as assertion inferring. To deal with this issue, we used well-tested programs and state-of-the-art tools to generate extensive pools of tests (Evosuite~\cite{DBLP:conf/sigsoft/FraserA11} and Randoop~\cite{DBLP:conf/icse/PachecoLEB07}) as done by previous work \cite{DBLP:conf/sigsoft/TerragniJTP20, DBLP:conf/icse/MolinaPAF21, DBLP:conf/icse/MolinadA22}. This is also  a typical process followed in mutation testing studies~\cite{PapadakisHHJT16, KurtzAODKG16, JustKA17, 9677967}. To be more accurate, our underlying assumption is that the extensive pool of tests used in our experiments is a valid approximation of the program's test executions. 

\textit{Construct Validity}: 
Our assessment metrics, mutation filtered assertions inferred, ground truth assertions inferred, and incurred time during mutation analysis may not reflect the actual cost~/~benefit values. These metrics are intuitive, i.e., the inferred assertions are the output of assertion inference techniques, and the incurred time during mutation testing is the wall clock time these techniques invest in filtering assertions. Overall, we mitigate these threats by following suggestions from mutation testing and assertion inference literature, using state of the art tools, performing several simulations, and got consistent and stable results across our subjects.
\section{Conclusion}
\label{sec_conclusion}
We presented \our, a method that learns to select \aim (a small subset of mutants that is suitable for assertion inference) from given mutant sets. Our experiments on 40 subjects show that \our identified assertion inferring mutants with 0.58 MCC, 0.79 Precision, and 0.49 Recall. These predictions enable 42.29 times faster inference with minor effectiveness loss (12.49\% less assertions) compared to the use of all mutants. Similarly, \our's predictions infer 96.15\% of the total ground truth assertions, which is 40\% more than \sms and 5 times more than \rnd. Moreover, \our enables assertion inference technique SpecFuzzer to scale on all our large subjects (by inferring assertions where SpecFuzzer failed previously due to timeouts) in comparison to \rnd which failed to infer any assertion in 50\% of the large subjects. 

\bibliographystyle{plain}
\bibliography{Bibliography}

\begin{thebibliography}{10}

\bibitem{DBLP:journals/corr/AbadiABBCCCDDDG16}
Mart{\'{\i}}n Abadi, Ashish Agarwal, Paul Barham, Eugene Brevdo, Zhifeng Chen,
  Craig Citro, Gregory~S. Corrado, Andy Davis, Jeffrey Dean, Matthieu Devin,
  Sanjay Ghemawat, Ian~J. Goodfellow, Andrew Harp, Geoffrey Irving, Michael
  Isard, Yangqing Jia, Rafal J{\'{o}}zefowicz, Lukasz Kaiser, Manjunath Kudlur,
  Josh Levenberg, Dan Man{\'{e}}, Rajat Monga, Sherry Moore, Derek~Gordon
  Murray, Chris Olah, Mike Schuster, Jonathon Shlens, Benoit Steiner, Ilya
  Sutskever, Kunal Talwar, Paul~A. Tucker, Vincent Vanhoucke, Vijay Vasudevan,
  Fernanda~B. Vi{\'{e}}gas, Oriol Vinyals, Pete Warden, Martin Wattenberg,
  Martin Wicke, Yuan Yu, and Xiaoqiang Zheng.
\newblock Tensorflow: Large-scale machine learning on heterogeneous distributed
  systems.
\newblock {\em CoRR}, abs/1603.04467, 2016.

\bibitem{DBLP:conf/iclr/AlonBLY19}
Uri Alon, Shaked Brody, Omer Levy, and Eran Yahav.
\newblock code2seq: Generating sequences from structured representations of
  code.
\newblock In {\em 7th International Conference on Learning Representations,
  {ICLR} 2019, New Orleans, LA, USA, May 6-9, 2019}. OpenReview.net, 2019.

\bibitem{DBLP:conf/icst/AmmannDO14}
Paul Ammann, M{\'{a}}rcio~Eduardo Delamaro, and Jeff Offutt.
\newblock Establishing theoretical minimal sets of mutants.
\newblock In {\em Seventh {IEEE} International Conference on Software Testing,
  Verification and Validation, {ICST} 2014, March 31 2014-April 4, 2014,
  Cleveland, Ohio, {USA}}, pages 21--30. {IEEE} Computer Society, 2014.

\bibitem{DBLP:books/daglib/0020331}
Paul Ammann and Jeff Offutt.
\newblock {\em Introduction to Software Testing}.
\newblock Cambridge University Press, 2008.

\bibitem{DBLP:journals/corr/BahdanauCB14}
Dzmitry Bahdanau, Kyunghyun Cho, and Yoshua Bengio.
\newblock Neural machine translation by jointly learning to align and
  translate.
\newblock In Yoshua Bengio and Yann LeCun, editors, {\em 3rd International
  Conference on Learning Representations, {ICLR} 2015, San Diego, CA, USA, May
  7-9, 2015, Conference Track Proceedings}, 2015.

\bibitem{DBLP:conf/icassp/BahdanauCSBB16}
Dzmitry Bahdanau, Jan Chorowski, Dmitriy Serdyuk, Philemon Brakel, and Yoshua
  Bengio.
\newblock End-to-end attention-based large vocabulary speech recognition.
\newblock In {\em 2016 {IEEE} International Conference on Acoustics, Speech and
  Signal Processing, {ICASSP} 2016, Shanghai, China, March 20-25, 2016}, pages
  4945--4949. {IEEE}, 2016.

\bibitem{DBLP:journals/tse/BarrHMSY15}
Earl~T. Barr, Mark Harman, Phil McMinn, Muzammil Shahbaz, and Shin Yoo.
\newblock The oracle problem in software testing: {A} survey.
\newblock {\em {IEEE} Trans. Software Eng.}, 41(5):507--525, 2015.

\bibitem{DBLP:conf/issta/BlasiGKGEPC18}
Arianna Blasi, Alberto Goffi, Konstantin Kuznetsov, Alessandra Gorla,
  Michael~D. Ernst, Mauro Pezz{\`{e}}, and Sergio~Delgado Castellanos.
\newblock Translating code comments to procedure specifications.
\newblock In Frank Tip and Eric Bodden, editors, {\em Proceedings of the 27th
  {ACM} {SIGSOFT} International Symposium on Software Testing and Analysis,
  {ISSTA} 2018, Amsterdam, The Netherlands, July 16-21, 2018}, pages 242--253.
  {ACM}, 2018.

\bibitem{DBLP:journals/ml/Breiman01}
Leo Breiman.
\newblock Random forests.
\newblock {\em Mach. Learn.}, 45(1):5--32, 2001.

\bibitem{DBLP:journals/corr/BritzGLL17}
Denny Britz, Anna Goldie, Minh{-}Thang Luong, and Quoc~V. Le.
\newblock Massive exploration of neural machine translation architectures.
\newblock {\em CoRR}, abs/1703.03906, 2017.

\bibitem{ChekamPBTS20}
Thierry~Titcheu Chekam, Mike Papadakis, Tegawend{\'{e}}~F. Bissyand{\'{e}},
  Yves~Le Traon, and Koushik Sen.
\newblock Selecting fault revealing mutants.
\newblock {\em Empir. Softw. Eng.}, 25(1):434--487, 2020.

\bibitem{DBLP:conf/emnlp/ChoMGBBSB14}
Kyunghyun Cho, Bart van Merrienboer, {\c{C}}aglar G{\"{u}}l{\c{c}}ehre, Dzmitry
  Bahdanau, Fethi Bougares, Holger Schwenk, and Yoshua Bengio.
\newblock Learning phrase representations using {RNN} encoder-decoder for
  statistical machine translation.
\newblock In Alessandro Moschitti, Bo~Pang, and Walter Daelemans, editors, {\em
  Proceedings of the 2014 Conference on Empirical Methods in Natural Language
  Processing, {EMNLP} 2014, October 25-29, 2014, Doha, Qatar, {A} meeting of
  SIGDAT, a Special Interest Group of the {ACL}}, pages 1724--1734. {ACL},
  2014.

\bibitem{DBLP:journals/sigsoft/ClarkeR06}
Lori~A. Clarke and David~S. Rosenblum.
\newblock A historical perspective on runtime assertion checking in software
  development.
\newblock {\em {ACM} {SIGSOFT} Softw. Eng. Notes}, 31(3):25--37, 2006.

\bibitem{DBLP:conf/kbse/dAmorimPXME06}
Marcelo d'Amorim, Carlos Pacheco, Tao Xie, Darko Marinov, and Michael~D. Ernst.
\newblock An empirical comparison of automated generation and classification
  techniques for object-oriented unit testing.
\newblock In {\em 21st {IEEE/ACM} International Conference on Automated
  Software Engineering {(ASE} 2006), 18-22 September 2006, Tokyo, Japan}, pages
  59--68. {IEEE} Computer Society, 2006.

\bibitem{DBLP:conf/issta/DemskyEGMPR06}
Brian Demsky, Michael~D. Ernst, Philip~J. Guo, Stephen McCamant, Jeff~H.
  Perkins, and Martin~C. Rinard.
\newblock Inference and enforcement of data structure consistency
  specifications.
\newblock In Lori~L. Pollock and Mauro Pezz{\`{e}}, editors, {\em Proceedings
  of the {ACM/SIGSOFT} International Symposium on Software Testing and
  Analysis, {ISSTA} 2006, Portland, Maine, USA, July 17-20, 2006}, pages
  233--244. {ACM}, 2006.

\bibitem{DBLP:journals/scp/ErnstPGMPTX07}
Michael~D. Ernst, Jeff~H. Perkins, Philip~J. Guo, Stephen McCamant, Carlos
  Pacheco, Matthew~S. Tschantz, and Chen Xiao.
\newblock The daikon system for dynamic detection of likely invariants.
\newblock {\em Sci. Comput. Program.}, 69(1-3):35--45, 2007.

\bibitem{DBLP:conf/sigsoft/FraserA11}
Gordon Fraser and Andrea Arcuri.
\newblock Evosuite: automatic test suite generation for object-oriented
  software.
\newblock In Tibor Gyim{\'{o}}thy and Andreas Zeller, editors, {\em
  SIGSOFT/FSE'11 19th {ACM} {SIGSOFT} Symposium on the Foundations of Software
  Engineering {(FSE-19)} and ESEC'11: 13th European Software Engineering
  Conference (ESEC-13), Szeged, Hungary, September 5-9, 2011}, pages 416--419.
  {ACM}, 2011.

\bibitem{DBLP:journals/tse/FraserZ12}
Gordon Fraser and Andreas Zeller.
\newblock Mutation-driven generation of unit tests and oracles.
\newblock {\em {IEEE} Trans. Software Eng.}, 38(2):278--292, 2012.

\bibitem{DBLP:conf/issta/GaleottiRPF10}
Juan~P. Galeotti, Nicol{\'{a}}s Rosner, Carlos~L{\'{o}}pez Pombo, and
  Marcelo~F. Frias.
\newblock Analysis of invariants for efficient bounded verification.
\newblock In Paolo Tonella and Alessandro Orso, editors, {\em Proceedings of
  the Nineteenth International Symposium on Software Testing and Analysis,
  {ISSTA} 2010, Trento, Italy, July 12-16, 2010}, pages 25--36. {ACM}, 2010.

\bibitem{DBLP:journals/ese/GargDJCPT22}
Aayush Garg, Renzo Degiovanni, Matthieu Jimenez, Maxime Cordy, Mike Papadakis,
  and Yves~Le Traon.
\newblock Learning from what we know: How to perform vulnerability prediction
  using noisy historical data.
\newblock {\em Empir. Softw. Eng.}, 27(7):169, 2022.

\bibitem{9677967}
Aayush Garg, Milos Ojdanic, Renzo Degiovanni, Thierry~Titcheu Chekam, Mike
  Papadakis, and Yves Le~Traon.
\newblock Cerebro: Static subsuming mutant selection.
\newblock {\em IEEE Transactions on Software Engineering}, pages 1--1, 2022.

\bibitem{GongZYM17}
Dunwei Gong, Gongjie Zhang, Xiangjuan Yao, and Fanlin Meng.
\newblock Mutant reduction based on dominance relation for weak mutation
  testing.
\newblock {\em Information {\&} Software Technology}, 81:82--96, 2017.

\bibitem{GopinathAAJG16}
Rahul Gopinath, Mohammad~Amin Alipour, Iftekhar Ahmed, Carlos Jensen, and Alex
  Groce.
\newblock On the limits of mutation reduction strategies.
\newblock In Laura~K. Dillon, Willem Visser, and Laurie~A. Williams, editors,
  {\em Proceedings of the 38th International Conference on Software
  Engineering, {ICSE} 2016, Austin, TX, USA, May 14-22, 2016}, pages 511--522.
  {ACM}, 2016.

\bibitem{Jimenez2019realworld}
Matthieu Jimenez, Renaud Rwemalika, Mike Papadakis, Federica Sarro, Yves
  Le~Traon, and Mark Harman.
\newblock The importance of accounting for real-world labelling when predicting
  software vulnerabilities.
\newblock In {\em Proceedings of the 2019 27th ACM Joint Meeting on European
  Software Engineering Conference and Symposium on the Foundations of Software
  Engineering}, ESEC/FSE 2019, page 695–705, New York, NY, USA, 2019.
  Association for Computing Machinery.

\bibitem{JustKA17}
Ren{\'{e}} Just, Bob Kurtz, and Paul Ammann.
\newblock Inferring mutant utility from program context.
\newblock In {\em Proceedings of the 26th {ACM} {SIGSOFT} International
  Symposium on Software Testing and Analysis, Santa Barbara, CA, USA, July 10 -
  14, 2017}, pages 284--294, 2017.

\bibitem{DBLP:conf/kbse/JustSK11}
Ren{\'{e}} Just, Franz Schweiggert, and Gregory~M. Kapfhammer.
\newblock {MAJOR:} an efficient and extensible tool for mutation analysis in a
  java compiler.
\newblock In Perry Alexander, Corina~S. Pasareanu, and John~G. Hosking,
  editors, {\em 26th {IEEE/ACM} International Conference on Automated Software
  Engineering {(ASE} 2011), Lawrence, KS, USA, November 6-10, 2011}, pages
  612--615. {IEEE} Computer Society, 2011.

\bibitem{DBLP:conf/emnlp/KalchbrennerB13}
Nal Kalchbrenner and Phil Blunsom.
\newblock Recurrent continuous translation models.
\newblock In {\em Proceedings of the 2013 Conference on Empirical Methods in
  Natural Language Processing, {EMNLP} 2013, 18-21 October 2013, Grand Hyatt
  Seattle, Seattle, Washington, USA, {A} meeting of SIGDAT, a Special Interest
  Group of the {ACL}}, pages 1700--1709. {ACL}, 2013.

\bibitem{DBLP:conf/apsec/KintisPM10}
Marinos Kintis, Mike Papadakis, and Nicos Malevris.
\newblock Evaluating mutation testing alternatives: {A} collateral experiment.
\newblock In Jun Han and Tran~Dan Thu, editors, {\em 17th Asia Pacific Software
  Engineering Conference, {APSEC} 2010, Sydney, Australia, November 30 -
  December 3, 2010}, pages 300--309. {IEEE} Computer Society, 2010.

\bibitem{KurtzAODKG16}
Bob Kurtz, Paul Ammann, Jeff Offutt, M{\'{a}}rcio~Eduardo Delamaro, Mariet
  Kurtz, and Nida G{\"{o}}k{\c{c}}e.
\newblock Analyzing the validity of selective mutation with dominator mutants.
\newblock In Thomas Zimmermann, Jane Cleland{-}Huang, and Zhendong Su, editors,
  {\em Proceedings of the 24th {ACM} {SIGSOFT} International Symposium on
  Foundations of Software Engineering, {FSE} 2016, Seattle, WA, USA, November
  13-18, 2016}, pages 571--582. {ACM}, 2016.

\bibitem{DBLP:journals/scp/LeavensCCRC05}
Gary~T. Leavens, Yoonsik Cheon, Curtis Clifton, Clyde Ruby, and David~R. Cok.
\newblock How the design of {JML} accommodates both runtime assertion checking
  and formal verification.
\newblock {\em Sci. Comput. Program.}, 55(1-3):185--208, 2005.

\bibitem{DBLP:conf/oopsla/LogozzoB12}
Francesco Logozzo and Thomas Ball.
\newblock Modular and verified automatic program repair.
\newblock In Gary~T. Leavens and Matthew~B. Dwyer, editors, {\em Proceedings of
  the 27th Annual {ACM} {SIGPLAN} Conference on Object-Oriented Programming,
  Systems, Languages, and Applications, {OOPSLA} 2012, part of {SPLASH} 2012,
  Tucson, AZ, USA, October 21-25, 2012}, pages 133--146. {ACM}, 2012.

\bibitem{MarcozziBKPPC18}
Micha{\"{e}}l Marcozzi, S{\'{e}}bastien Bardin, Nikolai Kosmatov, Mike
  Papadakis, Virgile Prevosto, and Lo{\"{\i}}c Correnson.
\newblock Time to clean your test objectives.
\newblock In Michel Chaudron, Ivica Crnkovic, Marsha Chechik, and Mark Harman,
  editors, {\em Proceedings of the 40th International Conference on Software
  Engineering, {ICSE} 2018, Gothenburg, Sweden, May 27 - June 03, 2018}, pages
  456--467. {ACM}, 2018.

\bibitem{DBLP:books/ph/Meyer97}
Bertrand Meyer.
\newblock {\em Object-Oriented Software Construction, 2nd Edition}.
\newblock Prentice-Hall, 1997.

\bibitem{DBLP:conf/icse/MolinadA22}
Facundo Molina, Marcelo d'Amorim, and Nazareno Aguirre.
\newblock Fuzzing class specifications.
\newblock In {\em 44th {IEEE/ACM} 44th International Conference on Software
  Engineering, {ICSE} 2022, Pittsburgh, PA, USA, May 25-27, 2022}, pages
  1008--1020. {ACM}, 2022.

\bibitem{DBLP:conf/icse/MolinaPAF21}
Facundo Molina, Pablo Ponzio, Nazareno Aguirre, and Marcelo~F. Frias.
\newblock Evospex: An evolutionary algorithm for learning postconditions.
\newblock In {\em 43rd {IEEE/ACM} International Conference on Software
  Engineering, {ICSE} 2021, Madrid, Spain, 22-30 May 2021}, pages 1223--1235.
  {IEEE}, 2021.

\bibitem{DBLP:conf/icse/PachecoLEB07}
Carlos Pacheco, Shuvendu~K. Lahiri, Michael~D. Ernst, and Thomas Ball.
\newblock Feedback-directed random test generation.
\newblock In {\em 29th International Conference on Software Engineering {(ICSE}
  2007), Minneapolis, MN, USA, May 20-26, 2007}, pages 75--84. {IEEE} Computer
  Society, 2007.

\bibitem{PapadakisHHJT16}
Mike Papadakis, Christopher Henard, Mark Harman, Yue Jia, and Yves~Le Traon.
\newblock Threats to the validity of mutation-based test assessment.
\newblock In Andreas Zeller and Abhik Roychoudhury, editors, {\em Proceedings
  of the 25th International Symposium on Software Testing and Analysis, {ISSTA}
  2016, Saarbr{\"{u}}cken, Germany, July 18-20, 2016}, pages 354--365. {ACM},
  2016.

\bibitem{DBLP:journals/ac/PapadakisK00TH19}
Mike Papadakis, Marinos Kintis, Jie Zhang, Yue Jia, Yves~Le Traon, and Mark
  Harman.
\newblock Chapter six - mutation testing advances: An analysis and survey.
\newblock {\em Adv. Comput.}, 112:275--378, 2019.

\bibitem{DBLP:conf/sosp/PerkinsKLABCPSSSWZER09}
Jeff~H. Perkins, Sunghun Kim, Samuel Larsen, Saman~P. Amarasinghe, Jonathan
  Bachrach, Michael Carbin, Carlos Pacheco, Frank Sherwood, Stelios Sidiroglou,
  Gregory~T. Sullivan, Weng{-}Fai Wong, Yoav Zibin, Michael~D. Ernst, and
  Martin~C. Rinard.
\newblock Automatically patching errors in deployed software.
\newblock In Jeanna~Neefe Matthews and Thomas~E. Anderson, editors, {\em
  Proceedings of the 22nd {ACM} Symposium on Operating Systems Principles 2009,
  {SOSP} 2009, Big Sky, Montana, USA, October 11-14, 2009}, pages 87--102.
  {ACM}, 2009.

\bibitem{Pinto2020vocabulary}
Gustavo Pinto, Breno Miranda, Supun Dissanayake, Marcelo d'Amorim, Christoph
  Treude, and Antonia Bertolino.
\newblock What is the vocabulary of flaky tests?
\newblock In {\em Proceedings of the 17th International Conference on Mining
  Software Repositories}, MSR '20, page 492–502, New York, NY, USA, 2020.
  Association for Computing Machinery.

\bibitem{DBLP:journals/tse/ShepperdBH14}
Martin~J. Shepperd, David Bowes, and Tracy Hall.
\newblock Researcher bias: The use of machine learning in software defect
  prediction.
\newblock {\em {IEEE} Trans. Software Eng.}, 40(6):603--616, 2014.

\bibitem{DBLP:journals/jaiscr/ShewalkarNL19}
Apeksha Shewalkar, Deepika Nyavanandi, and Simone~A. Ludwig.
\newblock Performance evaluation of deep neural networks applied to speech
  recognition: Rnn, {LSTM} and {GRU}.
\newblock {\em J. Artif. Intell. Soft Comput. Res.}, 9(4):235--245, 2019.

\bibitem{DBLP:conf/nips/SutskeverVL14}
Ilya Sutskever, Oriol Vinyals, and Quoc~V. Le.
\newblock Sequence to sequence learning with neural networks.
\newblock In Zoubin Ghahramani, Max Welling, Corinna Cortes, Neil~D. Lawrence,
  and Kilian~Q. Weinberger, editors, {\em Advances in Neural Information
  Processing Systems 27: Annual Conference on Neural Information Processing
  Systems 2014, December 8-13 2014, Montreal, Quebec, Canada}, pages
  3104--3112, 2014.

\bibitem{DBLP:conf/sigsoft/TerragniJTP20}
Valerio Terragni, Gunel Jahangirova, Paolo Tonella, and Mauro Pezz{\`{e}}.
\newblock Evolutionary improvement of assertion oracles.
\newblock In Prem Devanbu, Myra~B. Cohen, and Thomas Zimmermann, editors, {\em
  {ESEC/FSE} '20: 28th {ACM} Joint European Software Engineering Conference and
  Symposium on the Foundations of Software Engineering, Virtual Event, USA,
  November 8-13, 2020}, pages 1178--1189. {ACM}, 2020.

\bibitem{DBLP:conf/tap/TillmannH08}
Nikolai Tillmann and Jonathan de~Halleux.
\newblock Pex-white box test generation for .net.
\newblock In Bernhard Beckert and Reiner H{\"{a}}hnle, editors, {\em Tests and
  Proofs - 2nd International Conference, {TAP} 2008, Prato, Italy, April 9-11,
  2008. Proceedings}, volume 4966 of {\em Lecture Notes in Computer Science},
  pages 134--153. Springer, 2008.

\bibitem{DBLP:conf/icse/TufanoPWBP19}
Michele Tufano, Jevgenija Pantiuchina, Cody Watson, Gabriele Bavota, and Denys
  Poshyvanyk.
\newblock On learning meaningful code changes via neural machine translation.
\newblock In Joanne~M. Atlee, Tevfik Bultan, and Jon Whittle, editors, {\em
  Proceedings of the 41st International Conference on Software Engineering,
  {ICSE} 2019, Montreal, QC, Canada, May 25-31, 2019}, pages 25--36. {IEEE} /
  {ACM}, 2019.

\bibitem{DBLP:journals/tosem/TufanoWBPWP19}
Michele Tufano, Cody Watson, Gabriele Bavota, Massimiliano~Di Penta, Martin
  White, and Denys Poshyvanyk.
\newblock An empirical study on learning bug-fixing patches in the wild via
  neural machine translation.
\newblock {\em {ACM} Trans. Softw. Eng. Methodol.}, 28(4):19:1--19:29, 2019.

\bibitem{DBLP:conf/icsm/TufanoWBPWP19}
Michele Tufano, Cody Watson, Gabriele Bavota, Massimiliano~Di Penta, Martin
  White, and Denys Poshyvanyk.
\newblock Learning how to mutate source code from bug-fixes.
\newblock In {\em 2019 {IEEE} International Conference on Software Maintenance
  and Evolution, {ICSME} 2019, Cleveland, OH, USA, September 29 - October 4,
  2019}, pages 301--312. {IEEE}, 2019.

\bibitem{DBLP:conf/icse/WatsonTMBP20}
Cody Watson, Michele Tufano, Kevin Moran, Gabriele Bavota, and Denys
  Poshyvanyk.
\newblock On learning meaningful assert statements for unit test cases.
\newblock In Gregg Rothermel and Doo{-}Hwan Bae, editors, {\em {ICSE} '20: 42nd
  International Conference on Software Engineering, Seoul, South Korea, 27 June
  - 19 July, 2020}, pages 1398--1409. {ACM}, 2020.

\bibitem{DBLP:conf/ease/YaoS20}
Jingxiu Yao and Martin~J. Shepperd.
\newblock Assessing software defection prediction performance: why using the
  matthews correlation coefficient matters.
\newblock In Jingyue Li, Letizia Jaccheri, Torgeir Dings{\o}yr, and Ruzanna
  Chitchyan, editors, {\em {EASE} '20: Evaluation and Assessment in Software
  Engineering, Trondheim, Norway, April 15-17, 2020}, pages 120--129. {ACM},
  2020.

\bibitem{ZhangHHXM10}
Lu~Zhang, Shan{-}Shan Hou, Jun{-}Jue Hu, Tao Xie, and Hong Mei.
\newblock Is operator-based mutant selection superior to random mutant
  selection?
\newblock In Jeff Kramer, Judith Bishop, Premkumar~T. Devanbu, and
  Sebasti{\'{a}}n Uchitel, editors, {\em Proceedings of the 32nd {ACM/IEEE}
  International Conference on Software Engineering - Volume 1, {ICSE} 2010,
  Cape Town, South Africa, 1-8 May 2010}, pages 435--444. {ACM}, 2010.

\end{thebibliography}

\end{document}